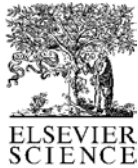
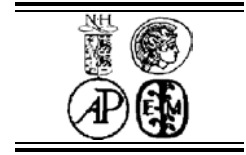

# Spin and Valence-Fluctuation Mediated Superconductivity in Pressurized Fe and CeCu$_2$(Si/Ge)$_2$

Didier Jaccard[*], Alexander T. Holmes

*University of Geneva, DPMC, 24 Quai E.-Ansermet, 1211 Genève 4, Switzerland*

**Abstract**

We review the evidence supporting valence-fluctuation mediated superconductivity in CeCu$_2$Si$_2$ and CeCu$_2$Ge$_2$, where $T_c$ reaches 2.4 K at high pressure. In these systems the valence and magnetic critical points, at $p_v$ and $p_c$ respectively, are well separated. Characteristic signatures associated with both are distinct. In contrast, the valence and spin fluctuation regions appear much closer in most Ce based compounds. Concerning $d$-transition metals, superconductivity in pure iron emerges in the pressure window 15-30 GPa with the onset of $T_c$ up to almost 3 K. All relevant observations point to unconventional superconductivity, likely mediated by ferromagnetic spin fluctuations. © 2001 Elsevier Science. All rights reserved

## 1. Introduction

The fascinating physical properties of SCES originate in the proximity of these systems to one or several phase transitions. To investigate these, pressure $p$ is often the ideal control parameter. However, due to experimental difficulties, most experiments have been done at ambient pressure using other control parameters like alloying (chemical pressure) or magnetic field, with unavoidable drawbacks such as atomic disorder in the case of alloying [1,2]. In recent years, the scenario of a 2$^{nd}$ order magnetic phase transition ending at zero temperature at a quantum critical point (QCP) has been promoted [3]. Magnetic fluctuations near the QCP, particularly those in reduced dimensions lead to deviations from Fermi liquid behaviour, and sometimes superconductivity. Differences between the alloying effect and the effect of pressure applied to a lattice clearly point to a major role for disorder, particularly in the ground state excitation resistivity relationship $\rho \propto T^n$ ($T \to 0$). Indeed, $n \approx 1$ was found for CeAu$_x$Cu$_{6-x}$ [4] and Ce(In$_{1-x}$Sn$_x$)$_3$ [5] at the critical concentration $x_c$, while $n \approx 1.5$ was observed at the critical pressure $p_c$ for CeCu$_5$Au [6] and CeIn$_3$ [7] respectively. The vanishing of magnetic order as a function of the control parameter was also found to be more sudden in the lattice case [4,6]. Moreover, in a recent analysis, little evidence was found for a 2$^{nd}$ order phase transition scenario [8]. In fact, most and perhaps all antiferromagnetic systems exhibit a 1$^{st}$ order like transition with a vertical magnetic line at $p_c$; in such a situation density fluctuations may be expected.

We present here two different cases for which the 1$^{st}$ order character of the transition is to some extent clear and the superconducting transition temperature $T_c$ is relatively high. In the first case of pure iron, the ferromagnetic transition is superimposed on a martensitic structural transformation. The existence of a QCP is not obvious, but in the pressure window 13-30 GPa the relevant properties of the normal and superconducting phases appear to be dominated by spin fluctuations. In the second case of CeCu$_2$(Si/Ge)$_2$ we discuss the valence transition located at a pressure $p_v$ which is well separated from the magnetic critical pressure $p_c$ in this system. Critical valence fluctuations cause various signatures to appear at $p_v$, and in particular $T_c$ is considerably enhanced. In most Ce compounds, the pressures $p_v$ and $p_c$ are much closer and an interplay of the associated phenomena occurs.

---
[*] Corresponding author. Tel.: +41-22-379-6366; fax: +41-22-379-6869; e-mail: didier.jaccard@physics.unige.ch.

## 2. Spin-mediated superconductivity in Fe

Soon after the discovery of superconductivity in pressurised iron by Shimizu et al. [9] we tried to clarify this challenging case. Some basic reasons were that iron is a simple element, which should be easier to approach theoretically, that the energy scale of the spin fluctuations may be very large, hence a high $T_c$, and that the topic fits our expertise well (!) [for experimental details see 10]. The $(p,T)$ phase diagram of solid iron is shown in Fig. 1b. Up to moderate $p$ and $T$, iron is in its ferromagnetic $\alpha$ phase, and has a *bcc* structure. Note that the *bcc* structure can be predicted by theory only if magnetic correlations are taken into account [11]. At relatively low pressure, the Curie point merges with the *bcc* $\alpha$ to *fcc* $\gamma$ frontier.

Above a pressure of around 13 GPa at 300 K, iron transforms into the *hcp* $\varepsilon$ phase, with a 5% volume reduction [13]. $\varepsilon$-iron is believed to be non-magnetic, Mössbauer effect measurements for example showing a fixed moment of less than 0.05 $\mu_B$ at 20 GPa in an ethanol/methanol pressure medium [14]. In Fig. 1b, the lines for the structural transitions are partly arbitrary because the latter are martensitic [15,16]. This means a large hysteresis and a subtle dependence on the pressure medium. The $\alpha$-$\varepsilon$ transition is non-diffusive and shear driven; it occurs by atomic plane slippage, with crystal orientation preserved. For a given pressure medium, the width of the transition is found to be about 10 times the pressure gradient $\delta p$ [17]. During the martensitic $\alpha$-$\varepsilon$ process proposed in ref. [16], there is an intermediate *fcc* arrangement, as if the $\gamma$ phase has a narrow dip between $\alpha$ and $\varepsilon$ phases, though there is likely to be a rich microstructure during the transition. Our high pressure resistivity experiments have been performed in a soft solid (steatite) medium in which $\delta p/p$ is about 5%, thus the $\alpha$-$\varepsilon$ transformation should be very gradual, extending over a large part of the superconducting pressure range.

At first sight, the $\alpha$-$\varepsilon$ transition is clearly seen in the $p$ dependence of the resistivity $\rho$ of iron at ambient as well as very low temperature as shown in Fig. 1a. Indeed, a steep and pronounced increase in $\rho(p)$ is observed around 13 GPa. This is followed by a broad maximum and a slow decrease above 16 GPa as $p$ is increased further. Initially, at low pressure, $\rho$ decreases very slowly and linearly with $p$. According to Mössbauer effect experiments [17], the $\alpha$-$\varepsilon$ transition should start at $p < 10$ GPa in a steatite medium. Thus the sharp kink in $\rho(p)$ likely corresponds to the pressure at which long range ferromagnetic order is lost, which may occur well after the beginning of $\alpha$-$\varepsilon$ transformation first appears. Moreover a comparison with the resistivity of metastable $\gamma$ iron [18] shows that the $\rho(p)$ increase at room temperature is unlikely to be due to a change in the electron-phonon term, but rather to a large spin-wave contribution. At low $T$, the residual $\rho_0(p)$ can jump by at least a factor or ten around 13 GPa for a good sample (i.e. when $\rho_0(p=0)$ is low). It seems that with the loss of ferromagnetic long-range correlations, the effect of static and dynamic spin disorder is strongly magnified. On increasing the pressure further, one can extrapolate the residual resistivity back to its value in the $\alpha$ phase at a pressure just above 30 GPa, which corresponds to the disappearance of superconductivity.

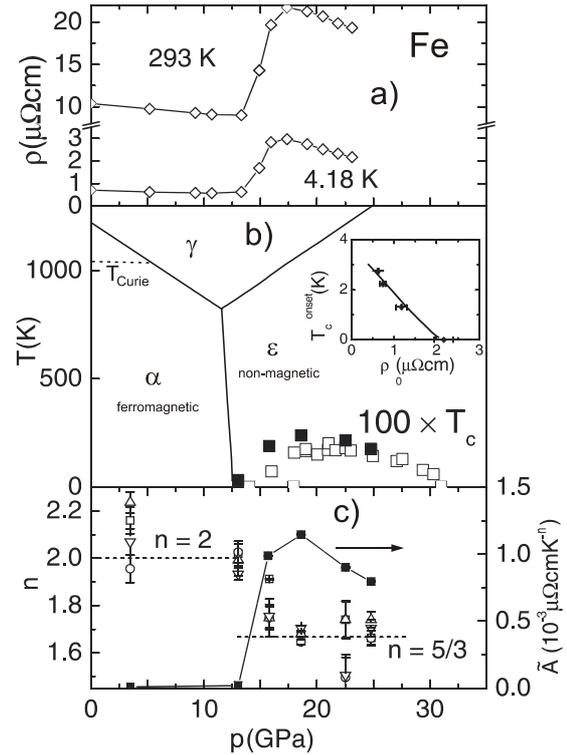

Fig.1. Plotted as a function of pressure are (a) the resistivity $\rho(p)$ of Fe at 293 K and 4.2 K, (b) the $(p,T)$ phase diagram of solid Fe showing the superconducting region (open squares from Ref. [9]) with $T_c^{onset}$ vs. $\rho_0$ in inset and (c) the exponent $n$ and coefficient $\tilde{A}$ in the relationship $\rho = \rho_0 + \tilde{A}T^n$, with different symbols referring to different samples.

According to Shimizu et al. [9] only very pure Fe can superconduct, so we restricted our investigations to samples with 99.99% purity or higher. We have confirmed the shape of the superconducting region (see Fig. 1b) for samples from three different sources [19]. The emergence of $T_c$ (from $T_c = 0$) just at the vanishing of ferromagnetism and the small pressure range of the superconducting domain have already been taken as signs of a non-phonon pairing mechanism [20,21]. Moreover, our results show that for samples with impurity levels below 100 ppm, the crucial parameter is not the chemical purity but the metallurgical state of the sample. Indeed by manipulating the intrinsic sample disorder, either by rolling (cold work induces defects such as dislocations) or conversely by annealing, an anticorrelation between $T_c$ and the residual resistivity $\rho_0$ has been demonstrated for pressure close to $p(T_c^{max})$, as shown in the inset of Fig. 1b. The latter gives strong support for spin-triplet pairing, as in the case of $Sr_2RuO_4$ [22]. The electronic mean free path $l \propto \rho_0^{-1}$ has a threshold value around 10 nm for superconductivity; according to the critical field data [19], the superconducting coherence length $\xi$ appears to be smaller than $l$, i.e. the clean limit is required, supporting an unconventional explanation for the pairing mechanism.

There is hope of obtaining higher $T_c$'s in future as lower $\rho_0$ values at ambient pressure lead to a lower increase at the $\alpha$-$\varepsilon$ transformation; an extrapolation suggests a maximum around almost 4 K. Even for good samples (i.e. those with a high $T_c$ onset and low $\rho_0$) the resistive transition remains very broad, barely reaching $\rho = 0$ for $T \leq 0.5$ K. A complete transition required a low current density $j \leq 1$ A/cm$^2$, while no decrease of the $T_c$ onset was detected for $j = 10^3$ A/cm$^2$. Moreover, more complete transitions were seen on the high pressure side of the $T_c(p)$ maximum. These latter features seem to indicate the pressure-induced growth of superconducting regions with weak links just before collapsing near 30 GPa.

Other support for spin-mediated superconductivity in Fe is given by the normal state resistivity described by the relationship $\rho = \rho_0 + \tilde{A}T^n$. The $n$ and $\tilde{A}$ values versus pressure are shown in Fig. 1c. Just at the emergence of superconductivity, the exponent $n$ jumps from the Fermi liquid value $n = 2$ to very near the value $n = 5/3$, characteristic of a nearly ferromagnetic Fermi liquid. This behaviour ($\rho \propto T^{5/3}$) extends up to around 30 K (i.e. several times $T_c$) and persists at least down to 1 K if a magnetic field is applied in order to suppress superconductivity [19]. The exponent $n = 3/2$ expected for antiferromagnetic spin fluctuations is clearly disproved, despite all calculations of *hcp* iron predicting an AFM ground state [23]. The exponent $n$ locks to 5/3, recalling the case of MnSi [24] and no QCP is seen, but rather a quantum region. Perhaps this is due to the 1$^{st}$ order and martensitic character of the $\alpha$-$\varepsilon$ transition. Finally, the value of the $\tilde{A}$ coefficient appears to track $T_c$, implying that the same fluctuations responsible for the non-Fermi liquid behaviour in resistivity may also be responsible for the pairing interaction. The large $\tilde{A}$ coefficient was found to be independent of the magnetic field up to 8 T.

Around 19 GPa, coinciding with the highest $T_c$, a fit to $(\rho - \rho_0)/T^2$ shows an enhancement of the effective mass by a factor of about 6 in comparison with $\alpha$-Fe. The open experimental question is whether superconductivity is intrinsic to *hcp* $\varepsilon$ iron or closely related to the change of the local Fe environment across the martensitic transformation. Very recent transport experiments made on whiskers seem to indicate a narrower $T_c$ domain [25]. Measurement in more hydrostatic media should clarify the interplay of structural, magnetic and superconducting properties.

## 3. Valence fluctuation mediated superconductivity in $CeCu_2(Si/Ge)_2$

It has long been known that the valence of certain ions in a solid can depend on a control parameter such as pressure. In the case of homogeneous valence, this change can be smooth or abrupt, accompanied or not by a structural transition. A particular example is given by metallic cerium whose $(p,T)$ phase diagram shows a 1$^{st}$ order valence discontinuity line [26]. This line separates the $\gamma$-Ce with a $4f$ shell occupation $n_f = 1$ from the $\alpha$-Ce which $n_f \cong 0.9$. The valence transition is isostructural (Ce has an *fcc* structure) and the line has a critical end point in the vicinity of $p_{cr} = 2$ GPa and $T_{cr} = 600$ K. For Ce based compounds an analogous situation has never clearly been reported, with the exception of the particular case of CeNi [27]. In cases where $p_{cr}$ is positive, either $T_{cr}$ is very high [28] or $T_{cr}$ seems to be negative and only a crossover regime is accessible even at $T = 0$. There are however possible exceptions like $CeCu_2Si_2$ and $CeCu_2Ge_2$ for which $T_{cr}$ is likely positive although small. In such a situation, the associated low energy valence fluctuations can mediate superconductivity.

The phase diagram of $CeCu_2Ge_2$, established from resistivity measurements, is shown in Fig. 2 [29]. Data for $CeCu_2Si_2$ superimpose well with a pressure shift of about 10 GPa (the Si ionic radius being smaller than that of Ge) and a slightly expanded $p$ scale (upper scale) as the Ge compound is stiffer at high pressure. The bulk modulus of $CeCu_2Ge_2$ at 10 K is $B_0 \approx 130$ GPa and the compression is almost isotropic [30]. In Fig. 2, arrows mark the two critical pressures $p_c$ and $p_v$. The latter is defined by the intercept of the valence discontinuity line at $T = 0$. Note that according to the $CeCu_2Ge_2$ data the magnetic line could be vertical at $p_c$, in contrast to $CeCu_2(Si,Ge)_2$ alloys [31]. Around $p_c$, a consensus has developed that superconductivity is spin-mediated. Near $p_v$, several anomalies are seen, to be discussed below. Notably, the resistivity is linear in $T$, and the superconducting transition temperature $T_c$ is maximum, valence fluctuations providing the pairing mechanism. According to lattice measurements, the end point of the valence line lies somewhere between 10 and 300 K [31].

We note the following milestones in the investigation of the relationship between valence change and superconductivity. Already proposed 20 years ago [10,32], a role for valence fluctuations in the pressure enhancement of $T_c$ in $CeCu_2Si_2$ received strong additional experimental support in 1998 [29] and was put on a theoretical footing by Miyake the same year [33]. Recently, it was shown that the two superconducting regions around $p_c$ and $p_v$ can be split by a pair breaking effect in CeCu$_2$(Si,Ge)$_2$ alloys [31]. Valence fluctuation mediated superconductivity received further support from ac specific heat and transport measurements in an He pressure cell, along with new theoretical developments [34]. Lastly, Monthoux and Lonzarich have proposed a theory of density-fluctuation-mediated superconductivity, which encompassed valence fluctuations [35].

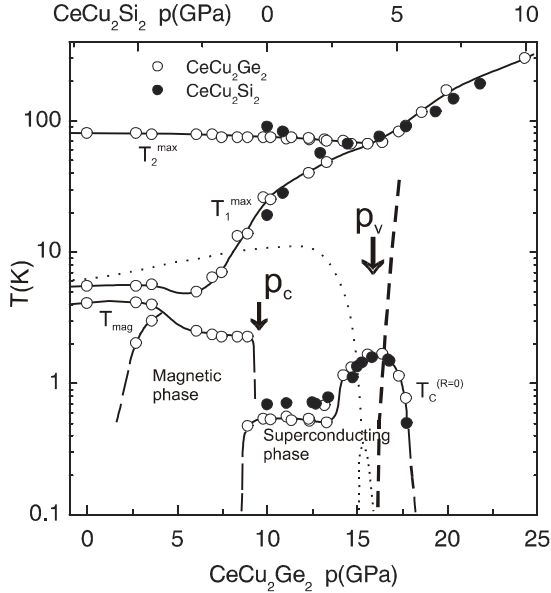

Fig. 2. *(p, T)* phase diagram of CeCu$_2$(Si/Ge)$_2$ showing the two critical pressures $p_c$ and $p_v$. Around $p_v$, superconductivity is mediated by valence fluctuations, the temperatures $T_1^{max}$ and $T_2^{max}$ merge and resistivity is *T* linear. The dashed line represents a hypothetical 1$^{st}$ order valence discontinuity. The dotted line represents the magnetic ordering temperature of CePd$_2$(Si/Ge)$_2$ and the narrow $T_c$ domain of the Si compound, plotted for matching $T_1^{max}$.

Miyake introduced an extra term into the periodic Anderson model, representing a Coulomb repulsion $U_{fc}$ between conduction *c*-, and *f*-electrons [36]. This term causes a rapid valence change when the $4f^1$ and $4f^0$ states become nearly degenerate under pressure. The associated fluctuations can scatter from *f* and conduction electrons and produce an attractive interaction in the *d*-wave channel. With increasing $U_{fc}$, the pressure induced valence change becomes more and more abrupt and the magnitude of the peak in $T_c$ increases. The major part of the scattering amplitude $\Gamma^{(0)}(q)$ is due to the scattering processes $(f,f) \leftrightarrow (f,c)$ in which the valence is changed directly. As a function of the momentum transfer *q*, $\Gamma^{(0)}(q)$ is roughly constant up to $3k_F/2$ reflecting the local nature of critical valence fluctuations. The residual resistivity $\rho_0$ was predicted to show a giant peak, scaling with the valence susceptibility $-(\partial n_f/\partial \varepsilon_f)_\mu$ where $\varepsilon_f$ is the atomic *f* level of the Ce ion and $\mu$ is the chemical potential. The $\rho_0$ peak is then expected to be located at $p_v$ [37]. This can be understood as the impurities nucleating a change of valence in the surrounding Ce ions, producing a greatly enhanced scattering cross section. Additionally, around $p_v$, the resistivity $\rho(T)$ should be *T*-linear in a large window and the $\gamma$ term of specific heat should exhibit a local maximum [34].

Let us now provide the list of the current experimental signatures associated with the valence transition in CeCu$_2$(Si/Ge)$_2$:

1) Cell volume [31] and $L_{III}$ x-ray absorption [38] measurements show a discontinuity around $p_v$ in CeCu$_2$Ge$_2$ and CeCu$_2$Si$_2$ respectively. The lack of anomaly at 300 K in the cell volume implies that the critical end points is located below room temperature. These experiments could be very sensitive to sample stoichiometry, so difficult to reproduce, the valence transition likely being weakly 1$^{st}$ order.

2) As shown in Fig. 3b, there are two regions where the predicted scaling $A \propto (T_1^{max})^{-2}$ is followed. As usual, $A$ is the coefficient of the $T^2$ resistivity law, always recovered at sufficiently low temperature and if necessary under magnetic field. $T_1^{max}$ is defined in the inset and assumed to be proportional to the energy scale given by $T_K$. In between these two regions, the compelling evidence for a valence transition is the abrupt drop in $A$ of over an order of magnitude, indicating the system is leaving the strongly correlated regime characterised by $n_f \approx 1$. This reminds us of the Kadowaki-Woods ratio $A/\gamma^2$ which crosses quickly from that of a strongly correlated class to a weakly correlated one [39]. The plot of Fig. 3b has been reproduced several times for both CeCu$_2$Si$_2$ and CeCu$_2$Ge$_2$, and has also been obtained in He pressure medium.

3) The bulk superconducting transition temperature $T_c$, as defined by the midpoint of the specific heat jump, is shown in Fig. 3a. Within the symbol size, this criterion for $T_c$ coincides with the completion of the resistive transition $T_c^{(R=0)}$, measured simultaneously on the same sample. Like in Fig. 3b, $T_c$ is plotted vs. $T_1^{max}$ in order to demonstrate the link with the valence change. It is indeed at the start of the $A$ drop that $T_c$ has a maximum, and superconductivity has disappeared by the point where the system has achieved its transition to the intermediate valence regime.

4) The residual resistivity $\rho_0$ has a huge peak at around the midpoint of the drop in $A$, i.e. at $p_v$. The differences in $\rho_0$ between samples at ambient pressure are vastly amplified. At $p_v$, $\rho_0$ can lie between 20 up to almost 200 $\mu\Omega$cm [10, 29, 34]. As a large part of $\rho_0$ comes from scattering processes which in fact contribute to the pair formation, the pair breaking effect of such huge $\rho_0$ is limited. Indeed $T_c^{(R=0)}$ is reduced by only a factor of two for the highest enhanced $\rho_0$ value.

5) The ac calorimetric experiment points to the occurrence of a maximum in the electronic specific heat coefficient $\gamma$ slightly below $p_v$, superimposed on a constant reduction with pressure [34]. This local maximum is supported by upper critical field measurements [10,25,40]. A close inspection of data from ref. [40] indicates the possibility that the $A$ coefficient of the resistivity also exhibits a local maximum at the same pressure.

6) Non-Fermi liquid behaviour occurs in the entire superconducting $p$ range [29,34]. One has $\rho \sim T^n$ with $n < 2$ from $T_c < T < T^*$. Notably, slightly below $p_v$, when $T_c$ is maximum, $n(p)$ takes its minimum value $n = 1$ and $T^*$ is maximum, which means that $\rho$ is $T$ linear up to about 25 K. There is a complication when $\rho_0$ is largest and $A$ reduced, as the impurity contribution has a negative $T$ variation [10].

7) As shown in Fig. 2, the temperatures $T_1^{max}$ and $T_2^{max}$, at which resistivity is maximum, merge around $p_v$. (For a definition of these temperatures see the inset of Fig. 3b). Basically the increase of $T_1^{max}$ scales the Kondo temperature $T_K$, while $T_2^{max}$ reflects the crystal field splitting $\Delta$. Further explanation of this characteristic is given in these proceedings [41].

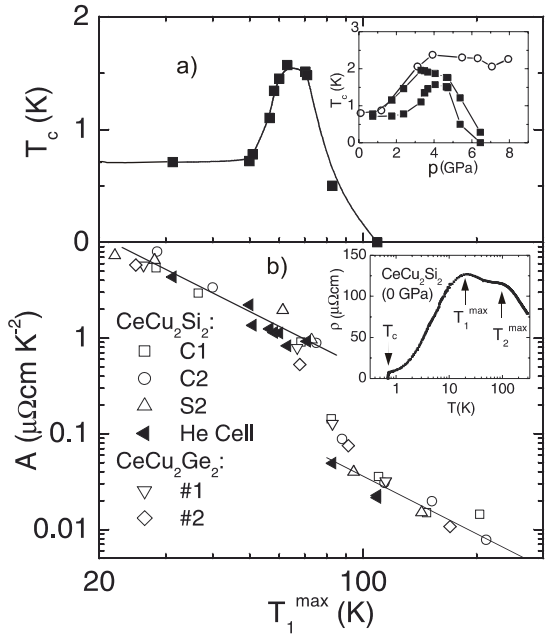

Fig. 3. Plotted against $T_1^{max}$ (defined in lower inset), a measure of the characteristic energy scale of the system, are (a) the bulk superconducting transition temperature and (b) the coefficient $A$ of $\rho \propto AT^2$ law of resisitivity. Note the straight lines where the $A \propto (T_1^{max})^{-2}$ scaling is followed. The upper inset shows $T_c^{bulk} \approx T_c^{(R=0)}$ and $T_c^{onset}$ obtained in the He cell (filled symbols), and $T_c^{onset}$ under strained conditions (open symbols).

Comparing experiment and the theory, it is noteworthy that almost all the features listed above, including superconductivity itself, follow directly from the valence fluctuation approach and the addition of a $U_{fc}$ term to the Hamiltonian. In particular the relative position of different anomalies are consistently reproduced on the pressure scale. Of course, other features have not been explained by valence fluctuation theory yet. They are, for example, the apparent increase around $p_v$ in the specific heat jump at $T_c$, and the fact that the upper critical field normalized to $T_c$, $H_{c2}(T/T_c)/T_c$ is weakly pressure dependent [40]. This latter aspect could be taken as a sign that the pairing mechanism is not so different at $p_c$ and $p_v$. Another mystery, already at experimental level, is the nature of the superconducting state between the onset and completion of the superconducting transition. The inset of Fig. 3a shows the transition onset $T_c^{onset}$ together with $T_c^{(R=0)}$ which nicely coincides with the bulk transition. As a function of pressure, $T_c^{onset}$ exhibits strange kinks similar to those reported in ref. [42] (and ascribed to topological changes of the Fermi surface). Values of $T_c^{onset}$ up to 2.4 K, with 90% of the resistance drop above 2.3 K have been previously reported, but $T_c^{(R=0)}$ never exceeded 1.8 K. This effect was never seen at low pressure. Contrary to the bulk $T_c$, the highest value of $T_c^{onset}$ is much more sensitive to the enhanced residual resistivity $\rho_0$, indicating a subtle pair breaking effect. So the question can be asked: is there any additional line in the $T_c(p)$ domain separating different superconducting phases?

A further point concerns the anisotropy. From experiments on $CePd_2Si_2$ [43] it was shown that the strain (due to the quasi-hydrostatic pressure conditions in Bridgman anvil cells) can have a dramatic effect on the superconducting and normal properties. In that compound, $p_c$ was shifted relative to the pure hydrostatic case, the $T_c(p)$ domain (especially if we consider $T_c^{onset}$ [44]) was greatly extended, and $T_c$ was raised by 40%. Recently, similar experiments have been done on $CeCu_2Si_2$ and the result for $T_c^{onset}$, shown in the inset of Fig. 3a, suggests that under strain part of the sample superconducts beyond 8 GPa. The observation that $T_c$ is very sensitive to (uniaxial) strain is consistent with the Monthoux & Lonzarich model [35]. These authors found that density-fluctuation mediated superconductivity (like for the magnetic scenario) is reinforced by more anisotropic structure.

The main experimental evidence that valence fluctuations are involved in the superconductivity of $CeCu_2(Si/Ge)_2$ was available in 1998, but the theoretical explanation proposed by Miyake was considered with scepticism. Problems with building a convincing case on the experimental side were due to on the one hand the very limited experimental possibilities at high pressure and on the other the uniqueness of $CeCu_2(Si/Ge)_2$. The latter can be addressed as follows: the crossing of the $T_1^{max}$ and $T_2^{max}$ temperatures (Fig. 2) is general and has been indeed observed in several systems. If this point is taken to be $p_v$, the comparison of many systems is possible.

Key ingredients for valence mediated superconductivity seem to be the relative strength of the magnetic RKKY, Kondo and crystal field (CF) interactions. In $CeCu_2(Si/Ge)_2$ the CF splitting $\Delta$ is relatively large while the magnetic RKKY interaction is weak. Thus at $p_c$, $T_K$ is low, i.e. the electronic specific heat term $\gamma$ is huge, and as $\Delta$ is large, $p_c \ll p_v$. Experimental signatures around $p_v$ associated with nearly critical valence fluctuations can be clearly identified. In comparison, for $CePd_2(Si/Ge)_2$, $T_1^{max}$ and $T_2^{max}$ are very similar to the Cu compound. But the magnetic interaction for some reason (role of Pd $d$-electrons?) is larger. Hence at $p_c$, a larger $T_K$ is required to balance the RKKY interaction, and so the $\gamma$ term is smaller. The magnetic phase vanishes close to $p_v$ where only a valence crossover is observed. There is no $\rho_0$ peak, but $\rho(T)$ is quasi linear over a broad $T$ range. The superconducting temperature $T_c$ is small and its pressure domain is narrow (in the hydrostatic case). According to Ref. [35], both spin mediated and valence mediated mechanisms could coexist, although the former is more robust.

The behaviour of $CePd_2(Si/Ge)_2$ is sketched in Fig. 2 by the dotted line (for data see [45]). In the case of $CeRh_2Si_2$, the magnetic interaction is stronger still and at $p_c$ the $\gamma$ term is lower still. The magnetic line is strongly $1^{st}$ order at $p_c$ and the superconducting domain shrinks with increasing sample quality [46]. Just above $p_c$ in $CePd_2Si_2$ one finds approximately $A \propto (T_I^{max})^{-4}$ [43], a much slower drop compared to the Cu compound. Heavy fermions like $CeAl_3$ or $CeCu_6$ have a low $T_K$ but moderate CF splitting $\Delta$. No trace of superconductivity was detected around $p_v$. Systems in which the magnetic and valence mechanisms could cooperate belong to the $CeTIn_5$ family, where linear resistivity, a $\rho_0$ peak, a maximum of $T_c(p)$ at $p > p_c$ and extended $T_c(p)$ domains have been observed [47]. Lastly, we note that the pressure $p_v$ corresponds to the "critical concentration" identified in a large number of Ce intermetallic alloys, analysed by their magnetic, thermal, transport, structural, and spectroscopic properties [48].

In any case the situation $p_c \ll p_v$ is rare among Ce compounds. Is there any hope with other rare earth, actinide or even $d$ systems?

## 4. Conclusion

In certain pressure intervals, critical fluctuations are responsible for the superconducting pairing interaction and anomalous normal phases properties of very different systems like Fe and $CeCu_2Si_2$ or its analogue $CeCu_2Ge_2$. In pressurised Fe, the pairing is likely to be of a spin-triplet nature, mediated by ferromagnetic fluctuations. Around a new type of quantum critical point, enhanced superconductivity in $CeCu_2(Si/Ge)_2$ is mediated by valence fluctuations.

## Acknowledgements

We thank Y. Onuki and G. Behr for providing us part of the Fe samples.

## References

[1] A. Schröder et al., Nature 407 (2000) 351
[2] J. Custers et al., Nature 424 (2003) 524
[3] N. D. Mathur et al., Nature 394 (1998) 39
[4] H. v. Löhneysen et al., Acta Phys. Polonica 34 (2003) 707 ; H. v. Löhneysen, J. Magn. Magn. Mat. 200 (1999) 532
[5] J. Custers et al., Acta Phys. Pol. 34 (2003) 379
[6] H. Wilhelm et al., in *Proc. AIRAPT-17, Hawaii, 1999*, eds. M. Manghanani et al. ,Universities Press, Hyderabad, 2000, p. 697. (cond-mat/9908442)
[7] G. Knebel et al., Phys. Rev. B 65 (2002) 024425
[8] J. Flouquet, Prog. Low Temp. Phys., to be published
[9] K. Shimizu et al., Nature 412 (2001) 316
[10] D. Jaccard et al., Rev. High Pressure Sci. Techn. 7 (1998) 412
[11] R. E. Cohen, cond-mat/0301615
[12] P. W. Mirwald and G. C. Kennedy, J. Geophys. Res., 84 (1979) 656
[13] A. P. Jephcoat, J. Geophys. Res. 91 (1986) 4677
[14] S. Nasu et al., J. Phys.: Condens. Matter 14 (2002) 11167
[15] W. A. Basset and H. Wang, Science 238 (1987) 780
[16] F. M. Wang and R. Ingalls, Phys. Rev. B 57 (1998) 5647
[17] R. D. Taylor et al., J. Appl. Phys. 69 (1991) 6126
[18] U. Bohnenkamp et al., J. Appl. Phys. 92 (1992) 4402
[19] A. T. Holmes et al., J. Phys.: Condens. Matter 16 (2004) S1121
[20] I. I. Mazin et al., Phys. Rev. B 65 (2002) 100511
[21] T. Jarlborg, Physica C 385 (2003) 513
[22] A. P. Mackenzie et al., Phys. Rev. Lett. 80 (1998) 161 ; A. P. Mackenzie and Y. Maeno, Rev. Mod. Phys. 75 (2003) 657
[23] V. Thakor et al., Phys. Rev. B 67 (2003) 180405
[24] N. Doiron-Leyraud et al., Nature 425 (2003) 595
[25] A. T. Holmes, Thesis (2004), unpublished
[26] D. C. Koskenmaki and K. A. Gschneidner Jr., in Handbook on the Physics and Chemistry of Rare Earths, vol 1, p. 337, Elsevier, Amsterdam, 1978
[27] D. Gignoux et al., J. Magn. Magn. Mat. 70 (1987) 388
[28] N. Môri et al., Jpn. J. Appl. Phys. 32 (1993) 300
[29] D. Jaccard et al., Physica B 259-261 (1999) 1
[30] A. Onodora et al., Solid State Commun. 123 (2002) 113
[31] H. Q. Yuan et al., Science 302 (2003) 2104
[32] B. Bellarbi et al., Phys. Rev. B 30 (1984) 1182 ; D. Jaccard et al., J. Magn. Magn. Mat. 47-48 (1985) 23
[33] K. Miyake et al., Physica B 259-261 (1985) 676
[34] A. T. Holmes et al., Phys. Rev. B 69 (2004) 024508
[35] P. Monthoux and G. G. Lonzarich, Phys. Rev. B 69 (2004) 064517
[36] Y. Onishi and K. Miyake, J. Phys. Soc. Jpn. 69 (2000) 3955
[37] K. Miyake and H. Maebashi, J. Phys. Soc. Jpn. 71 (2002) 1007
[38] J. Roehler et al., J. Magn. Magn. Mat. 76-77 (1988) 340
[39] K. Miyake et al., Solid State Commun. 71 (1989) 1149
[40] E. Vargoz et al., Solid State Commun. 106 (1998) 631
[41] Y. Nishida et al., this Proceedings
[42] F. Thomas et al., J. Phys.: Condens. Matter 8 (1996) L51
[43] A. Demuer et al., J. Phys.: Condens. Matter 14 (2002) L 529
[44] S. Raymond et al., Solid State Commun. 112 (1999) 617
[45] H. Wilhelm and D. Jaccard, Phys. Rev. B 66 (2002) 064428
[46] S. Araki et al., Acta Phys. Polonica 34 (2003) 439
[47] T. Muramatsu et al., J. Phys. Soc. Jpn. 70 (2001) 3362
[48] J. G. Sereni, Physica B 215 (1995) 273